\documentclass[12pt]{article}
\usepackage{graphicx,amsmath,amsfonts,amssymb,amscd,color,amsthm}
\newtheoremstyle{kai}
{3pt} {3pt} {} {} {\bfseries} {.} {.5em} {}
\def\EquationsBySection{\def\theequation
{\thesection.\arabic{equation}}%
\@addtoreset{equation}{section}}

\newcommand\old[1]{}
\newcommand{\pend}{\hfill \thicklines \framebox(6.6,6.6)[l]{}}
\renewenvironment{proof}{\noindent {\it  Proof.} \rm}{\pend}
\newtheorem{theorem}{Theorem}[section]
\newtheorem{lemma}{Lemma}[section]

\EquationsBySection \makeatother

\begin{document}
\pagestyle{plain}

\title
{\bf Markov chain-based stability analysis of growing networks}

\author{
 Zhenting Hou$^{1}$, Jinying Tong$^{1}${\thanks{Corresponding
 author,
  Email address: tongjy1120@yahoo.com.}, Dinghua Shi$^{1,2}$}
 \\  $^{1}$School of Mathematics, Central South University, \\Changsha, Hunan, 410075, China\\
   $^{2}$Department of Mathematics, Shanghai University,\\ Shanghai, 200444, China}

\date{\today}
\maketitle
\begin{abstract}{\rm
From the perspective of probability,  the stability of growing
network is studied in the present paper. Using the DMS model as an
example, we establish a relation between the growing network and
 Markov process. Based on the concept and technique of
first-passage probability in Markov theory, we provide a rigorous
proof for existence of the steady-state degree distribution,
mathematically re-deriving the exact formula of the distribution.
The approach based on Markov chain theory is universal and performs
well in a large class of growing networks.

  }

\end{abstract}
\noindent
\section{Introduction}
Networks are all around us, and we are ourselves, as individuals,
the units of a network of social relationships of different kinds
and, as biological systems, the delicate result of a network of
biochemical reactions. Networks can be tangible objects in the
Euclidean space, such as  the Internet, highways or subway systems,
and neural networks. Or they can be entities defined in an abstract
space, such as networks of acquaintances or collaborations between
individuals.\\
 \indent In the investigation of various complex networks, the steady-state
degree distributions of networks are always the main concerns
because they characterize the fundamental topological properties of
the underlying networks. The usual case in Science until a few years
ago was  homogeneous networks. Homogeneity in the interaction
structure means that almost all nodes are topologically equivalent,
like in regular lattices or in random graphs. However, when the
scientists approached the study of real networks from the available
databases, it was found that most of the real networks display power
law shaped degree distribution $p(k)\sim Ak^{-\gamma}$. Such
networks have been named scale-free networks, which was first
proposed by  Barab$\acute{a}$si and Albert$^{[1]}$ in 1999 (called
BA model hereafter).\\
 \indent The BA model is %a model of network growth inspired to the
%formation of the World Wide Web and is
starting with a small number ($m_0$) of nodes and based on two basic
ingredients: growth and preferential attachment. The probability
that a link of new node will connect to an existing node $i$ is
proportional to the degree of node $i$, i.e.,
%\begin{eqnarray}
$\Pi_{j\rightarrow i}=\frac{k_{i}}{\sum_{j}k_{j}}$ (how getting
started).
%\end{eqnarray}
 The BA model has been solved in the
mean-field approximation$^{[2]}$. There, to derive the following
dynamic equation:
\begin{eqnarray}
\frac{\partial k_{i}}{\partial t}=m\Pi(k_{i})=\frac{k_{i}}{2t},
\end{eqnarray}
it was assumed that the probability for an existing vertex to
receive a new connection from the new vertex is exactly equal to
$m\Pi(k_{i})$ (it is called $m\Pi$-hypothesis in [3]).%, which is
%simultaneously proportional to both the degree $q_{i}(t)$ of the
%existing vertex $i$ and the number of the new edges that the new
%vertex brings in,
%at time $t$.\\\indent
 %and, exactly, by means of rate equation[5] and
%master equation[1] approaches. In all the  works related to the BA
%model, the $m\Pi$-hypothesis plays a fundamental  role.  Assuming
%that the steady-state degree distribution exists,  the model
%produces a degree distribution $p(k)\sim k^{-\gamma}$, with an
%exponent$\gamma=3$.

 Since the ground-breaking papers by Barab$\acute{a}$si
and Albert$^{[1,2]}$ on scale-free networks, the interest on large
scale, growing  complex networks has soared. In addition to analytic
and numerical studies of the model itself, many authors have
proposed many modified and generalized models$^{[4,5,6]}$ such that
the model can fit real networks  well.
%\indent In [5], Krapivsky and Redner have considered a directed
%version of the BA model. They replaced the degree $q_{i}(t)$ of
%vertex i at time t by the total number $N_{k}(t)$ of degree-k
%vertexes over the whole network at time t, thereby obtained its
%rate equation
%\begin{eqnarray*}
%\frac{dN_{k}(t)}{dt}=m\frac{(k-1)N_{k-1}(t)-kN_{k}(t)}{\sum_{k}kN_{k}(t)}+\delta_{km},
%\end{eqnarray*}
%where $\delta_{km}$ accounts for new vertexes bringing in new
%edges. In their study, the $m\Pi-hypothesis$ was adopted in the
%derivations. Assuming that the steady-state degree distribution
%exists, using the law of large numbers
%($\frac{N_{k}(t)}{t}\rightarrow p(k)$ as $t\rightarrow\infty$),
%they showed that the difference equation of p(k) had an analytic
%solution
% \begin{eqnarray*}
%p(k)=\frac{4}{k(k+1)(k+2)},
%\end{eqnarray*}
% for the BA model with $m=1$. They yet pointed out that only the linear preferential attachment scheme
% could lead to the scale-free structure and any nonlinear one would
% not.\\

 Especially, S.N. Dorogovtsev, J.F.F. Mendes and A.N. Samukhin$^{[4]}$  proposed a general
attractiveness model(also called the DMS model in [5]).\\
\indent {\it At each time step a new site appears.
 Simultaneously, $m$ new directed links coming out from
non-specified sites are introduced. Let the connectivity $q_{s}$ be
the number of incoming links to a site $s$, i.e., to a site added at
time $s$. The probability that a new link points to a given site s
is proportional to the following characteristic of the site:
$A_{s}=A+q_{s}$, thereafter called its attractiveness. All sites are
born with some initial attractiveness $A\geq0$, but
afterwards it increases because of the $q_{s}$ term.}\\
\indent {\it It is convenient to assume that initially ($t=1$) we
have one site with m incoming links, when $t\geq 2$ sites are born
with zero connectivity (i.e., without incoming links).}\\
\indent {\it Note that one may allow multiple links, i.e., the
connectivity of a given site may increase simultaneously by more
than one. $\cdots$ The probability that a new link is connected with
the site s equals $\frac{A_{s}}{A_{\sum}}$, where $A_{\sum}=(m+A)s$.
The probability for the site s to receive exactly $l$ new links of
the $m$ injected is
$P_{s}^{(ml)}=(_{l}^{m})(\frac{A_{s}}{A_{\Sigma}})^{l}(1-\frac{A_{s}}{A_{\Sigma}})^{m-l}$.}

Dorogovtsev et al. first derived the master equation for the
distribution $P(q,s,t)$ of the connectivity $q$ of the site $s$.
Next, the connectivity distribution of the entire network is defined
by $P(q,t)=\frac{1}{t}\sum\limits_{i=1}^{t}P(q,i,t)$. Finally,
assuming that the limit $P(q)$ exists (note that actually an
additional assumption of
 $\lim\limits_{t\rightarrow\infty}t(P(q,t+1)-P(q,t))=0$ is also needed), they get the
following equation for the stationary connectivity distribution:
\begin{eqnarray}
(1+a)P(q)+(q+ma)P(q)-(q-1+ma)P(q-1)=(1+a)\delta_{q0},
\end{eqnarray}
where $a\triangleq A/m$. Then one may use the $Z$ transform of the
distribution function to obtain the following formulas:
\begin{eqnarray}
P(q)&=&(1+a)\frac{\Gamma[(m+1)a+1]}
{\Gamma(ma)}\frac{\Gamma(q+ma)}{\Gamma[q+2+(m+1)a]}.
\end{eqnarray}
%considered a
% network with
% linear preferential attachment(The Dorogovtsev-Mendes-Samukhin (DMS) model) of the form
% \begin{eqnarray}
%\Pi_{j\rightarrow i}=\frac{q_{i}+q_{0}}{\sum_{l}(q_{l}+q_{0})}.
%\end{eqnarray}
% This is a more general form than Eq.(1.1), because of the presence of
%the constant $q_{0}$, that plays the role of the node initial
%attractiveness. By assuming the existence of $p(k)$ ,  one gets a power law
%degree distribution with an exponent $\gamma = 2 + k_{0}/m$. \\
%\begin{remark}\\
\indent {\bf Remark:} The attractiveness model has overcome two
problems of the BA model, which were pointed out by
Bollob$\acute{a}$s in [6].
\\
\indent
 (1)The introduced parameter A, the initial
attractiveness, governs the probability for `young' or isolated
sites to get new links.
\\
\indent (2)Because a site may allow multiple links, one need no
$m\Pi$-hypothesis, though we have found an efficient way to realize
this hypothesis[2].\\
%\end{remark}
 \indent Shi et al.$^{[7,8]}$ established a corresponding relation between
growing networks and Markov chains, and proposed a computational
approach for network degree distributions. Consider the degree
$q_{i}(t)$ of node $i$ at time $t$. Based on the preferential
attachment mechanism of the BA model, the stochastic process
$q_{i}(t)$, $t=i, i+1, \cdots $ is a nonhomogeneous Markov chain.
Thus, the dynamics of a node from the time it joins the network is
described by a nonhomogeneous Markov chain and the whole network
(excluding the initial nodes) is completely described by a family of
Markov chains $\{q_{i}(t)\}$, $i=1, 2, \cdots  $, called network
Markov chain.\\
 \indent Our research is mainly motivated by   the following observation. \\
  (1)\;The existence of the steady-state
degree distribution  is  the basic issue in the research of network
degree distribution. However, from the discussion above,  we know
that it is unsolved in a considerable
amount of  models. \\
(2)\;There are all kinds of networks and  each  has its own evolving
mechanism, such as allowing  multiple links, loops or not, etc..
Therefore, it will be interesting if we can find a general way to
study them.\\
 \indent In the present paper, we will provide a general way to study the
steady-state degree distribution for the growing network.  To show
the feasibility and efficiency of our method, we concentrate
primarily on the DMS model. First, we will construct a network
Markov chain for the DMS model. Then, based on the concept and
techniques of first-passage probability proposed by Hou, et
al.$^{[3]}$, we provides a rigorous proof for the steady-state
degree distribution of the DMS model, and mathematically re-deriving
the exact analytic formulas of the distribution. But compared with
the paper[2],  the network Markov chain $\{q_{i}(t)\}$ here is a
general growth process which can jump $m$ states after a transition,
not pure-birth process being considered in [3]. Thus, we greatly
generalize the results obtained in Hou et al..

% \textit{The DMS Model[1].}
%\indent To provide a rigorous proof of the stability of the
%scale-free network for the DMS model, it is convenient to assume
%that the initially($t=1$) we have one site with $m$ incoming
%links-the resulting behavior at long time is independent of the
%initial condition.

\section{Stability analysis}
The attachment mechanism of the DMS model indicates that the future
evolution of the in-degree $q_{i}(t)$ as a process is independent of
 the past history, given its current state. We can observe that
the sequence $\{q_{i}(t),\ t=i,i+1,\cdots\}$ is a nonhomogeneous
Markov chain with the state space $\Omega=\{0,1,2,\cdots\}$. Thus,
the state transition probabilities of the Markov chain are given by
\begin{eqnarray}
p(q_{i}(t+1)=l|q_{i}(t)=q) &=&\left\{
\begin{array}{ll}
(^{m}_{j})[\frac{q+A}{(m+A)t}]^{j}[1-\frac{q+A}{(m+A)t}]^{m-j},\; l=q+j,0\leq j\leq m;\\
0,\;\;\;\;\;otherwise.
\end{array}
\right.
\end{eqnarray}
where $q=0,1,2,\cdots, m(t-i)$ and $i=2,3,\cdots$.

Let $P(q,i,t)=p\{q_{i}(t)=q\}$ denote the probability of vertex $i$
having in-degree $q$ at time $t$, which has initial distribution as
follows
\begin{eqnarray*}
P(q,1,1)=p\{q_{1}(1)=q\}=\delta_{q,m},~~~~~~~~~~~~~\\
P(q,i,i)=p\{q_{i}(i)=q\}=\delta_{q,0},\;\; for\;\;
 i\geq2.
\end{eqnarray*}
 Define the degree distribution of the whole network by the average
value of probabilities of the vertex degrees
\begin{eqnarray}
P(q,t)\triangleq\frac{1}{t}\sum\limits_{i=1}^{t}P(q,i,t)=\frac{1}{t}P(q,1,t)+\frac{t-1}{t}
\frac{1}{t-1}\sum\limits_{i=2}^{t}P(q,i,t).
\end{eqnarray}
 Next, denote $\bar{P}(q,t)\triangleq
\frac{1}{t-1}\sum\limits_{i=2}^{t}P(q,i,t)$. We only need to prove
that $\lim\limits_{t\rightarrow\infty}\bar{P}(q,t)$ exists, which
will imply that
$\lim\limits_{t\rightarrow\infty}P(q,t)=\lim\limits_{t\rightarrow\infty}\bar{P}(q,t)$
exists.\\
\indent Denote the first-passage probability of the Markov chain by
$f(q,i,s)=P\{q_{i}(s)=q,q_{i}(l)\neq q,l=1,2,\cdots,s-1\}$. %, when
%$s\geq T$,  where $c$ is a positive integer and $T$ is defined as
%follows:
%\begin{eqnarray*}
%T=
%\end{eqnarray*}
%Otherwise,  $f(k,i,s)=0$. First,
Relationships between the first-passage probability and the
probability of vertex degrees are established.
\begin{lemma}
For $q>0,$
\begin{eqnarray}
f(q,i,s)=\sum_{j=1}^{\min(m,j)}
(_{j}^{m})\left[\frac{q-j+A}{(m+A)(s-1)}\right]^{j}\left[1-\frac{q-j+A}{(m+A)(s-1)}\right]^{m-j}P(q-j,i,s-1),~\\
P(q,i,t)=\sum_{s=t_0}^{t}f(q,i,s)\prod\limits_{j=s}^{t-1}\left[1-\frac{q+A}{(m+A)j}\right]^m,
~t_0=\left\{
\begin{array}{ll}
i+[q/m],&[q/m]=q/m;\\
i+[q/m]+1,&[q/m]\neq~q/m.
\end{array}
\right.
\end{eqnarray}
\end{lemma}
\begin{proof}
First, consider Eq.(2.3). According to the construction of the
attractiveness model, the in-degree of a vertex is always
nondecreasing, and increasing at most by $m$ each time. Thus, it
follows from Eq.(2.1) that
\begin{eqnarray*}
f(q,i,s)&=&P\{q_{i}(s)=q,q_{i}(l)\neq q,l=1,2,\cdots,s-1\}\\
&=&P\{q_{i}(s)=q,q_{i}(s-1)= q-1,q_{i}(l)\neq
q,l=1,2,\cdots,s-2\}+\cdots\\
&&+P\{q_{i}(s)=q,q_{i}(s-1)= q-m\geq 0,q_{i}(l)\neq
q,l=1,2,\cdots,s-2\}\\
&=&P\{q_{i}(s)=q,q_{i}(s-1)= q-1\}+\cdots+P\{q_{i}(s)=q,q_{i}(s-1)=
q-m\}\\
&=&\sum_{j=1}^{\min(m,j)}P(q-j,i,s-1)p\{q_{i}(s)=q|q_{i}(s-1)= q-j\}\\
 &=&\sum_{j=1}^{\min(m,j)}
(_{j}^{m})\left[\frac{q-j+A}{(m+A)(s-1)}\right]^{j}\left[1-\frac{q-j+A}{(m+A)(s-1)}\right]^{m-j}P(q-j,i,s-1).
 \end{eqnarray*}
\indent Second, observe that the earliest time for the in-degree of
vertex $i$ to reach $q$ is at step $t_0$, and the latest time to do
so is at step $t$. After this vertex degree becomes $q$, it will not
increase any more. Thus, Eq.(2.4) is proved.
\end{proof}

\begin{lemma}(Stolz-Ces$\acute{a}$ro Theorem)
Let $x_{n}$ and $y_{n}$ be two sequences of real numbers. If $y_{n}$
is positive, strictly increasing and unbounded and the following
limit exists:
$$\lim\limits_{n\rightarrow\infty}\frac{x_{n+1}-x_{n}}{y_{n+1}-y_{n}}=l.$$
Then the limit $\lim\limits_{n\rightarrow\infty}\frac{x_{n}}{y_{n}}$
also exists and it is equal to $l$.
\end{lemma}
\begin{proof}
This is a classical result, see[9].
\end{proof}
\begin{lemma}
For the probability $P(0,t)$ defined in (2.2),
$\lim\limits_{t\rightarrow\infty}P(0,t)$ exists and is independent
of the initial network; moreover,
\begin{eqnarray*}
P(0)\triangleq
\lim\limits_{t\rightarrow\infty}P(0,t)=\frac{m+A}{m+A+mA}.
\end{eqnarray*}
\end{lemma}

\begin{proof}
By the definition of network degree and $P(0,t+1,t+1)=1$, together
with $P(0,i,t+1)=P(0,i,t)[1-\frac{A}{(m+A)t}]^m$,  it follows that
\begin{eqnarray*}
\bar{P}(0,t+1)&=&\frac{1}{t}\sum\limits_{i=2}^{t+1}P(0,i,t+1)\\
&=&\frac{1}{t}[\sum\limits_{i=2}^{t}P(0,i,t+1)+P(0,t+1,t+1)]\\
&=&\frac{1}{t}\sum\limits_{i=2}^{t}P(0,i,t)[1-\frac{A}{(m+A)t}]^m+\frac{1}{t}\\
&=&\frac{t-1}{t}\bar{P}(0,t)[1-\frac{A}{(m+A)t}]^m+\frac{1}{t}.
\end{eqnarray*}
Then, by iteration, we have
\begin{eqnarray*}
\bar{P}(0,t)&=&\prod\limits_{i=2}^{t-1}\frac{i-1}{i}[1-\frac{A}{(m+A)i}]^m[\bar{P}(0,2)+\sum\limits_{l=2}^{t-1}
\frac{\frac{1}{l-1}}{\prod\limits_{j=2}^{l}[1-\frac{A}{(m+A)j}]^m\cdot\frac{j-1}{j}}]\\
&=&\frac{1}{t-1}\prod\limits_{i=2}^{t-1}[1-\frac{A}{(m+A)i}]^m
[1+\sum\limits_{l=2}^{t-1}\prod\limits_{j=2}^{l}[1-\frac{A}{(m+A)j}]^{-m}].
\end{eqnarray*}
Next, let
\begin{eqnarray*}
&&x_{n}=1+\sum\limits_{l=2}^{n-1}\prod\limits_{j=2}^{l}[1-\frac{A}{(m+A)j}]^{-m},\\
&&y_{n}=(n-1)\prod\limits_{i=2}^{n-1}[1-\frac{A}{(m+A)i}]^{-m}.
\end{eqnarray*}
Thus, it follows that
\begin{eqnarray*}
&&x_{n+1}-x_{n}=\prod\limits_{i=2}^{n}[1-\frac{A}{(m+A)i}]^{-m},\\
&&y_{n+1}-y_{n}=\prod\limits_{i=2}^{n-1}[1-\frac{A}{(m+A)i}]^{-m}[n[1-\frac{A}{(m+A)n}]^{-m}-(n-1)].
\end{eqnarray*}
\indent Since $y_{n}>0$ and $y_{n+1}-y_{n}>0$, $\{y_{n}\}$ is a
strictly monotone increasing nonnegative sequence, hence,
$y_{n}\rightarrow\infty$. Moreover,
\begin{eqnarray*}
\frac{x_{n+1}-x_{n}}{y_{n+1}-y_{n}}&=&\frac{[1-\frac{A}{(m+A)n}]^{-m}}
{n[1-\frac{A}{(m+A)n}]^{-m}-(n-1)}\\
&\rightarrow&\frac{m+A}{m+A+mA}\;(n\rightarrow\infty).
\end{eqnarray*}
From Lemma 2.2, one has
\begin{eqnarray*}
P(0)\triangleq \lim\limits_{t\rightarrow\infty}P(0,t)
=\lim\limits_{n\rightarrow\infty}\frac{x_{n}}{y_{n}}
=\lim\limits_{n\rightarrow\infty}\frac{x_{n+1}-x_{n}}{y_{n+1}-y_{n}}=\frac{m+A}{m+A+mA}.
\end{eqnarray*}
This completes the proof.
\end{proof}

\begin{lemma}
For any positive integer $q$, if
$\lim\limits_{t\rightarrow\infty}P(q-1,t)$ exists, then
$\lim\limits_{t\rightarrow\infty}P(q,t)$ also exists and, moreover,
\begin{eqnarray}
P(q)\triangleq
\lim\limits_{t\rightarrow\infty}P(q,t)=\frac{m(q+A-1)}{m(q+A+1)+A}P(q-1,t).
\end{eqnarray}
\end{lemma}

\begin{proof}
For any positive integer $q$, using Eq.(2.1) and (2.3) we see that
\begin{eqnarray*}
\bar{P}(q,t)&=&\frac{1}{t-1}\sum\limits_{i=2}^{t}P(q,i,t)\\
&=&\frac{1}{t-1}\sum\limits_{i=2}^{t}\sum\limits_{s=t_{0}}^{t}f(q,i,s)\prod\limits_{j=s}^{t-1}
[1-\frac{q+A}{(m+A)j}]^{m}\\
&=&\frac{1}{t-1}\sum\limits_{i=2}^{t}\sum\limits_{s=[\frac{q}{m}]+i}^{t}f(q,i,s)\prod\limits_{j=s}^{t-1}
[1-\frac{q+A}{(m+A)j}]^{m}.
\end{eqnarray*}
For simplicity, let
\begin{eqnarray*}
A_{j}=(_{j}^{m})[\frac{q-j+A}{(m+A)(s-1)}]^{j}[1-\frac{q-j+A}{(m+A)(s-1)}]^{m-j},\;\;1\leq
j\leq m,
\end{eqnarray*}
and
\begin{eqnarray*}
\bar{P}^{l}(q,t)=\frac{1}{t-1}\sum\limits_{i=2}^{t}\sum\limits_{s=[\frac{q}{m}]+i}^{t}A_{l}
P(q-l,i,s-1)\prod\limits_{j=s}^{t-1}
[1-\frac{q+A}{(m+A)j}]^{m},\;\;1\leq l\leq m.
\end{eqnarray*}
 Then, by Eq.(2.2), we see that
\begin{eqnarray*}
\bar{P}(q,t)&=&\frac{1}{t-1}\sum\limits_{i=2}^{t}P(q,i,t)\\
&=&\frac{1}{t-1}\sum\limits_{i=2}^{t}\sum\limits_{s=[\frac{q}{m}]+i}^{t}[A_{1}P(q-1,i,s-1)
+A_{2}P(q-2,i,s-1)\\
&&+\cdots+A_{m}P(q-m,i,s-1)]\prod\limits_{j=s}^{t-1}
[1-\frac{q+A}{(m+A)j}]^{m}\\
&=&\bar{P}^{1}(q,t)+\bar{P}^{2}(k,t)+\cdots+\bar{P}^{m}(q,t).
\end{eqnarray*}

(1)Consider the case that the degree of vertex increases only by 1
 at each step:
\begin{eqnarray*}
&&\bar{P}^{1}(q,t)\\
&=&\frac{1}{t-1}\sum\limits_{i=2}^{t}\sum\limits_{s=[\frac{q}{m}]+i}^{t}mP(q-1,i,s-1)\frac{q+A-1}{(m+A)(s-1)}
[1-\frac{q+A-1}{(m+A)(s-1)}]^{m-1}\prod\limits_{j=s}^{t-1}
[1-\frac{q+A}{(m+A)j}]^{m}\\
&=&\frac{m}{t-1}\sum\limits_{s=[\frac{q}{m}]+2}^{t}\sum\limits_{i=2}^{s-[\frac{q}{m}]}P(q-1,i,s-1)\frac{q+A-1}{(m+A)(s-1)}
[1-\frac{q+A-1}{(m+A)(s-1)}]^{m-1}\prod\limits_{j=s}^{t-1}[1-\frac{q+A}{(m+A)j}]^{m}\\
&=&\frac{m}{t-1}\sum\limits_{s=[\frac{q}{m}]+2}^{t}\bar{P}(q-1,s-1)\frac{q+A-1}{m+A}
[1-\frac{q+A-1}{(m+A)(s-1)}]^{m-1}\prod\limits_{j=s}^{t-1}[1-\frac{q+A}{(m+A)j}]^{m}\\
&=&\frac{m}{t-1}\frac{q+A-1}{m+A}\prod\limits_{j=[\frac{q}{m}]+2}^{t-1}[1-\frac{q+A}{(m+A)j}]^{m}
\{\bar{P}(q-1,[\frac{q}{m}]+1)[1-\frac{q+A-1}{(m+A)([\frac{q}{m}]+1)}]^{m-1}\\
&&+\sum\limits_{s=[\frac{q}{m}]+1}^{t-1}\bar{P}(q-1,s)[1-\frac{q+A}{(m+A)s}]^{m-1}
\prod\limits_{j=[\frac{q}{m}]+1}^{s}[1-\frac{q+A}{(m+A)j}]^{-m}\}.
\end{eqnarray*}
Let
\begin{eqnarray*}
x_{n}&=&\bar{P}(q-1,[\frac{q}{m}]+1)[1-\frac{q+A-1}{(m+A)([\frac{q}{m}]+1)}]^{m-1}\\
&&+\sum\limits_{s=[\frac{q}{m}]+1}^{n-1}\bar{P}(q-1,s)[1-\frac{q+A}{(m+A)s}]^{m-1}
\prod\limits_{j=[\frac{q}{m}]+1}^{s}[1-\frac{q+A}{(m+A)j}]^{-m},\\
y_{n}&=&\frac{n-1}{m}\frac{m+A}{q+A-1}\prod\limits_{j=[\frac{q}{m}]+1}^{n-1}[1-\frac{q+A}{(m+A)j}]^{-m}.
\end{eqnarray*}
Obviously,
\begin{eqnarray*}
&&x_{n+1}-x_{n}=\bar{P}(q-1,n)[1-\frac{q+A-1}{(m+A)n}]^{m-1}\prod\limits_{j=[\frac{q}{m}]}^{n}[1-\frac{q+A}{(m+A)j}]^{-m},\\
&&y_{n+1}-y_{n}=\frac{(m+A)}{m(q+A-1)}\prod\limits_{j=[\frac{q}{m}]}^{n-1}
[1-\frac{q+A}{(m+A)j}]^{-m}[n[1-\frac{q+A}{(m+A)n}]^{-m}-(n-1)]>0.
\end{eqnarray*}
\indent Since $\{y_{n}\}$ is a strictly monotone increasing
nonnegative sequence, hence $y_{n}\rightarrow\infty$. Also, by
assumption,
\begin{eqnarray*}
\frac{x_{n+1}-x_{n}}{y_{n+1}-y_{n}}
&=&\frac{\bar{P}(q-1,n)[1-\frac{q+A-1}{(m+A)n}]^{m-1}[1-\frac{q+A}{(m+A)n}]^{-m}}{\frac{(m+A)}{m(q+A-1)}[n
[1-\frac{q+A}{(m+A)n}]^{-m}-(n-1)]}\\
&=&\frac{m(q+A-1)}{m(q+A+1)+A}\bar{P}(q-1,n).
\end{eqnarray*}
From Lemma 2.2, one has
\begin{eqnarray*}
\lim\limits_{t\rightarrow\infty}\bar{P}^{1}(q,t)
=\lim\limits_{n\rightarrow\infty}\frac{x_{n}}{y_{n}}
=\lim\limits_{n\rightarrow\infty}\frac{x_{n+1}-x_{n}}{y_{n+1}-y_{n}}=\frac{m(q+A-1)}{m(q+A+1)+A}\bar{P}(q-1,n).
\end{eqnarray*}
(2) Consider the case that the degree of
vertexes increases by 2 simultaneously at each step:
\begin{eqnarray*}
&&\bar{P}^{2}(q,t)\\
&=&\frac{1}{t-1}\sum\limits_{i=2}^{t}\sum\limits_{s=[\frac{q}{m}]+i}^{t}P(q-2,i,s-1)
( _{2}^{m})[\frac{q+A-2}{(m+A)(s-1)}]^{2}
[1-\frac{q+A-2}{(m+A)(s-1)}]^{m-2}\prod\limits_{j=s}^{t-1}
[1-\frac{q+A}{(m+A)j}]^{m}\\
&=&\frac{(_{2}^{m})}{t-1}\sum\limits_{s=[\frac{q}{m}]+2}^{t}\bar{P}(q-2,s-1)\frac{(q+A-2)^2}{(m+A)^{2}(s-1)}
[1-\frac{q+A-2}{(m+A)(s-1)}]^{m-2}\prod\limits_{j=s}^{t-1}[1-\frac{q+A}{(m+A)j}]^{m}\\
&=&\frac{(_{2}^{m})(q+A-2)^2}{(m+A)^{2}(t-1)}\prod\limits_{j=[\frac{q}{m}]+2}^{t-1}[1-\frac{q+A}{(m+A)j}]^{m}
\{\bar{P}(q-2,[\frac{q}{m}]+1)\frac{1}{[\frac{q}{m}]+1}[1-\frac{q+A-2}{(m+A)([\frac{q}{m}]+1)}]^{m-2}\\
&&+\sum\limits_{s=[\frac{q}{m}]+1}^{t-1}\bar{P}(q-2,s)\frac{1}{s}[1-\frac{q+A-2}{(m+A)s}]^{m-2}
\prod\limits_{j=[\frac{q}{m}]+1}^{s}[1-\frac{q+A}{(m+A)j}]^{-m}\}.
\end{eqnarray*}
Let
\begin{eqnarray*}
x_{n}&=&\bar{P}(q-2,[\frac{q}{m}]+1)\frac{1}{[\frac{q}{m}]+1}[1-\frac{q+A-2}{(m+A)([\frac{q}{m}]+1)}]^{m-2}\\
&&+\sum\limits_{s=[\frac{q}{m}]+1}^{n-1}\bar{P}(q-2,s)\frac{1}{s}[1-\frac{q+A-2}{(m+A)s}]^{m-2}
\prod\limits_{j+1}^{s}[1-\frac{q+A}{(m+A)j}]^{-m}],\\
y_{n}&=&\frac{(n-1)(m+A)^2}{(_{2}^{m})(q+A-2)^2}\prod\limits_{j=[\frac{q}{m}]+2}^{n-1}[1-\frac{q+A}{(m+A)j}]^{-m}.
\end{eqnarray*}
Obviously,
\begin{eqnarray*}
&&x_{n+1}-x_{n}=\bar{P}(k-2,n)\frac{1}{n}[1-\frac{q+A-2}{(m+A)n}]^{m-2}
\prod\limits_{j=[\frac{q}{m}]+1}^{n}[1-\frac{q+A}{(m+A)j}]^{-m},\\
&&y_{n+1}-y_{n}=\frac{(m+A)^2}{(_{2}^{m})(q+A-2)^2}\prod\limits_{j=[\frac{q}{m}]+2}^{n-1}[1-\frac{q+A}{(m+A)j}]^{-m}
[n[1-\frac{q+A}{(m+A)n}]^{-m}-(n-1)]>0.
\end{eqnarray*}
\indent Since $\{y_{n}\}$ is a strictly monotone increasing
nonnegative sequence, hence $y_{n}\rightarrow\infty$. Also, by
assumption,
\begin{eqnarray*}
\frac{x_{n+1}-x_{n}}{y_{n+1}-y_{n}}&=&\frac{\bar{P}(q-2,n)\frac{1}{n}[1-\frac{q+A}{(m+A)n}]^{-m}
[1-\frac{q+A-2}{(m+A)n}]^{m-2}}
{\frac{(m+A)^2}{(_{2}^{m})(q+A-2)^2}[n[1-\frac{q+A}{(m+A)n}]^{-m}-(n-1)]}\\
&=&O(\frac{1}{n})\rightarrow0\; \; \;\;(when
\;\;\;n\rightarrow\infty).
\end{eqnarray*}
From Lemma 2.2, one has
\begin{eqnarray*}
\lim\limits_{t\rightarrow\infty}\bar{P}^{2}(q,t)
=\lim\limits_{n\rightarrow\infty}\frac{x_{n}}{y_{n}}
=\lim\limits_{n\rightarrow\infty}\frac{x_{n+1}-x_{n}}{y_{n+1}-y_{n}}=0.
\end{eqnarray*}
Similarly, for $2<i\leq m$, we can prove
$$\bar{P}^{i}(q,t)=o(\frac{1}{t})\rightarrow0 \;\;\;(when\;\; t\rightarrow\infty).$$
Hence,
\begin{eqnarray*}
\lim_{t\rightarrow\infty}P(q,t)&=&\frac{m(q+A-1)}{m(q+A+1)+A}P(q-1).
\end{eqnarray*}
\end{proof}
\begin{theorem}
For any positive integer $q$, the steady-state degree distribution
P(q) of the DMS model exists, and is given by
 \begin{eqnarray}
P(q)&=&\frac{m+A}{m}\frac{\Gamma(A+\frac{A}{m}+1)}
{\Gamma(A)}\frac{\Gamma(q+A)}{\Gamma(q+A+2+\frac{A}{m})}\\
 &\sim&\frac{m+A}{m}\frac{\Gamma(A+\frac{A}{m}+1)}
{\Gamma(A)}q^{-(2+\frac{A}{m})}\;\;(for\;large\;q).
\end{eqnarray}
\end{theorem}
\begin{proof}
By induction, applying Lemma 2.3 and 2.4, we easily see that the
steady-state degree distribution of the DMS model exists.  Following
from Eq.(2.4) by iteration till $k=1$,
\begin{eqnarray*}
\lim_{t\rightarrow\infty}P(q,t)&=&\frac{m(q+A-1)}{m(q+A+1)+A}P(q-1)\\
&=&\frac{\Gamma(A+\frac{A}{m}+2)}{\Gamma(A)}\frac{\Gamma(q+A)}{\Gamma(q+A+2+\frac{A}{m})}P(0)\\
&=&\frac{m+A}{m}\frac{\Gamma(A+\frac{A}{m}+1)}
{\Gamma(A)}\frac{\Gamma(q+A)}{\Gamma(q+A+2+\frac{A}{m})}.
\end{eqnarray*}

For large $q$ values the above formula simplifies further, because
in that case
$\lim\limits_{q\rightarrow\infty}\frac{\Gamma(q)}{\Gamma(q+p)}=q^{-p}$.
 Thus, for large $q$, we have
 \begin{eqnarray*}
P(q)\sim \frac{m+A}{m}\frac{\Gamma(A+\frac{A}{m}+1)}
{\Gamma(A)}q^{-(2+\frac{A}{m})}.
\end{eqnarray*}
\end{proof}

Specially, if one sets $A=m$ and  every new site is the source of
all the $m$ new links, in this case, the model that we consider is
equivalent to the BA model, except permitting multiple links.  From
Eq.(2.4), we get by iteration
\begin{eqnarray*}
P(q)=\lim_{t\rightarrow\infty}P(q,t)&=&\frac{2m(m+1)}{(q+m)(q+m+1)(q+m+2)}\sim
2m(m+1)q^{-3}.
\end{eqnarray*}

\section{Conclusion}

(1) For general growing networks,  we can first establish the
corresponding Markov chain, and then prove the steady-state degree
distribution similarly.

 (2) From the proof above, we can conclude
that multiple edges do not affect the steady-state degree
distribution. Because for long times, the probability to receive
simultaneously more than one link of the $m$ is vanishing.

(3)  The process $q_{i}(t)$ considered in Hou, et al$^{[3]}$ is a
pure birth Markov chain, and the corresponding transition
probability matrix is
\begin{eqnarray}
 \left[
 \begin{array}{ccccccc}
 p_{mm}& p_{m(m+1)}&        &         &         &  &      \\
       &  p_{(m+1)(m+1)}& p_{(m+1)(m+2)}&         &         &  &      \\
       &        &\ddots&\ddots    &         &  &       \\
        &       &      &  p_{(m+t-i)(m+t-i)}& p_{(m+t-i)(m+t-i+1)}  &  &       \\
        &       &      &          &1        &0 &       \\
        &       &      &          &          &\ddots &
 \end{array}
  \right].
\end{eqnarray}
where $p_{lk}\triangleq p\{q_{i}(t+1)=k|q_{i}(t)=l\}$. However, in
the present paper,  what we study is a general growing process. The
corresponding transition probability matrix is
\begin{eqnarray}
 \left[
 \begin{array}{ccccccc}
 p_{00}& p_{01}&  \cdots    &   p_{0m}      &         &  &      \\
       &  p_{11}& p_{12}&     \cdots    &   p_{1(m+1)}      &  &      \\
       &        &\ddots&\ddots    &  \ddots       &  &       \\
        &       &      &  p_{[m(t-i)][m(t-i)]}& p_{[m(t-i)][m(t-i)+1]}  &  \cdots&   p_{[m(t-i)][2m(t-i)]}    \\
        &       &      &          &1        &0 &    \cdots   \\
        &       &      &        &          &\ddots &\\
 \end{array}
  \right].
\end{eqnarray}
 And when $p_{kl}=0,\;\; for \;\;k>l+1$ and $p_{kl}=0\;\;for \;\;k>m+t-i$ in (3.2), namely the vertex can only gain one link at a step, we
gain the results in [3]. Hence,  the results obtained in [3] are
generalized in this paper. More generally, we provide a general way
for the proof of the steady-state degree distribution of growing
network, whether allowing multiple links, loops or not .

\end{document}